\documentclass[showpacs,bibnotes,prl,twocolumn]{revtex4}%
\usepackage{amsmath}
\usepackage{graphicx}
\usepackage{amsfonts}
\usepackage{amssymb}%
\begin{document}
\title{Universal quantum computation in deocoherence-free subspace with neutral atoms}
\author{Peng Xue$^{1}$, and Yun-Feng Xiao$^{2}$}
\affiliation{$^{1}$Institute of Quantum Optics and Quantum Information of the Austrian
Academy of Science, A-6020 Innsbruck, Austria}
\affiliation{$^{2}$Key Laboratory of Quantum Information, University of Science and
Technology of China, Hefei 230026, China}
\date{\today}

\begin{abstract}
We show how realistic cavity-assisted interaction between neutral atoms and
coherent optical pulses, and measurement techniques, combined with optical
transportation of atoms, allow for a universal set of quantum gates acting on
decoherence-free subspace (DFS) in deterministic way. The logical qubits are
immunized to the dominant source of decoherece-----dephasing; while, the
influences of additional errors are shown by numerical simulations. We analyze
the performance and stability of all required operations and emphasize that
all techniques are feasible with current experimental technology.

\end{abstract}

\maketitle

\textit{Introduction.}-----Manipulation of atoms in microscopic traps is one
of the major highlights of the extraordinary progress experienced by atomic,
molecular and optical (AMO) physics over the past few years, and has led to
important successes in the implementation of quantum information processing
\cite{Monroe}. Hence, several implementations of neutral atoms quantum
computing, exploiting various trapping methods and entangling interactions,
have been proposed \cite{Zoller1,Zoller2,Turchette,Duan1,Xiao}. Nevertheless,
the experimental requirements with these approaches turn out to be very
challenging, such as, a large number of atoms each of which is strongly
coupled with cavity mode, individually addressing, and localization to the
Lamb-Dicke limit.

A quantum memory stores information in superposition states, but interactions
between the quantum memory and its environments destroy the stored
information, so called-----decoherence. Decoherence-free subspaces (DFSs) have
been proposed \cite{Guo} to protect fragile quantum information against the
detrimental effects of decoherence. There have been a lot of theoretical
researches for achieving fault tolerant universal quantum computation in DFSs
\cite{Whaley}. Also significant experimental efforts have been made for
realization of such a decoherent-free quantum memory in different physical
systems \cite{NIST,Kielpinski,Kwiat}.

In this Letter, we present a scheme to realize a universal set of quantum
gates in deterministic way, acting on neutral atoms through cavity-assisted
interaction of coherent optical pulses in DFS, which from the beginning
immunizes our logical qubits against the dominant source of
decoherence-----collective dephasing. Our idea is at least two-fold. First, we
implement computation using specific physical mechanisms that allow for gates
in the encoded space without any overhead associated with encoded gates.
Second, in our construction the system never leaves the DFS during the entire
execution of gates, so that fault tolerance is natural and, in stark contrast
to the usual situation in quantum error correction, necessitates no extra
resources during the computation.\begin{figure}[ptb]
\includegraphics[width=8cm,height=4cm]{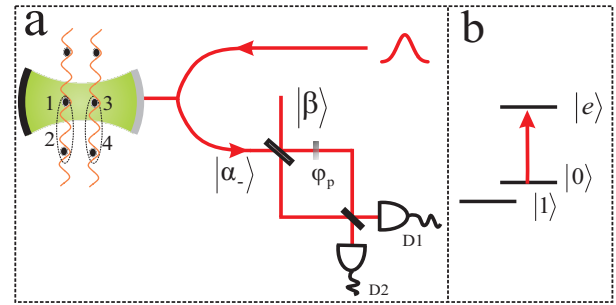}\newline\caption{(a) Schematic
setup for implementation of the logical CZ gate on two logical atomic qubits
in DFS through the cavity-assisted interaction. In order to verify the
projection, the scattering coherent optical pulses leaking out are detected by
the homodyne detectors after reflection. Here $\left\vert \beta\right\rangle $
is the state of local oscillator. (b) The relevant level structure of the atom
and the coupling configuration.}%
\end{figure}

Neutral atoms in our scheme are stored in transverse optical lattices and
translated into and outside of the cavity \cite{Sauer} for gate operations to
obviate the requirement for individual addressing, each of which has three
relevant levels as shown in Fig. 1. Atomic states $\left\vert 0\right\rangle $
and $\left\vert 1\right\rangle $ are two stable\ ground states. The atomic
transition from $\left\vert 0\right\rangle $ to excited level $\left\vert
e\right\rangle $\ is resonantly coupled to a cavity mode $a_{c}$. The state
$\left\vert 1\right\rangle $ is decoupled due to a large hyperfine splitting.
The coherent time of the superposition of the internal atomic states
$\tau_{co}$\ is with a magnitude of milliseconds \cite{Bonn}. There are two
dominant sources of decoherence: (i) photon loss during gate operations; (ii)
dephasing during the storage and transmission of the atoms in the optical
lattices. We will later show that in realistic setting gate errors due to
photon loss are characterized by the detailed numerical simulation which
demonstrates practicality of this scheme within the reach of the current
experimental technology. Furthermore, unlike single-photon detection
\cite{Duan1}, since a homodyne detection of the coherent state directly
measure the relative phase of the signal state, the photon losses only
decrease the signal to noise ration but not lead to a failure in the
measurements. Also we describe a specific encoding that allows suppression of
the error of type (ii) by considering a DFS by the states $\left\vert
0_{L}\right\rangle =\left\vert 01\right\rangle $ and $\left\vert
1_{L}\right\rangle =\left\vert 10\right\rangle $, which from the beginning
immunizes our logical qubits against the dominant source of
decoherence-----dephasing provided by stray fields, random variation of the
atom-cavity coupling rate and the instability of the optical lattice. We
denote the logical Bell states by $\left\vert \Phi_{L}^{\pm}\right\rangle
=\left(  \left\vert 0_{L}0_{L}\right\rangle \pm\left\vert 1_{L}1_{L}%
\right\rangle \right)  /\sqrt{2}=\left(  \left\vert 0101\right\rangle
\pm\left\vert 1010\right\rangle \right)  /\sqrt{2}$ and $\left\vert \Psi
_{L}^{\pm}\right\rangle =\left(  \left\vert 0_{L}1_{L}\right\rangle
\pm\left\vert 1_{L}0_{L}\right\rangle \right)  /\sqrt{2}=\left(  \left\vert
0110\right\rangle \pm\left\vert 1001\right\rangle \right)  /\sqrt{2}$, which
take the full advantage of these properties, suppressing phase noise. The
logical qubit decoheres only insofar as the dephasing fails to be collective.

\textit{Dynamical decoupling pulses and their application.}-----We briefly
review the decoupling technique \cite{Lidar} as it pertains to our problem.
Assume a phase noise term $\varepsilon\left(  t\right)  $ acts on the internal
states of atoms, characterized by a power spectrum $S\left(  \omega\right)  $
of integrated power $\left(  \tau_{co}\right)  ^{2}$ with a high frequency
cutoff at $\omega_{c}\ll1/\tau_{co}$. The action of $\varepsilon\left(
t\right)  $ can be represented by a stochastic evolution operator
$U_{x}\left(  t\right)  =e^{-i\int_{0}^{t}\varepsilon\left(  t^{\prime
}\right)  dt^{\prime}\sigma_{x}^{L}},$ where $\sigma_{x}^{L}$ is a Pauli
operator for the encoded subspace, which can be implemented simply by swapping
the two qubits. The pulse sequence [$\Delta t,U_{x},\Delta t,U_{x}$] gives a
reduced power spectrum $S_{DFS}\left(  \omega\right)  \propto S\left(
\omega\right)  \sin^{4}\left(  \Delta t\omega/2\right)  /\left(  \Delta
t\omega\right)  ^{2}$, where $\Delta t$ is free evolution time (cycle time).
For frequencies below $1/\Delta t,$ the bath-induced error rate is reduced by
a factor proportional to $\left(  \Delta t\omega\right)  ^{2}$.

The DFS also reduces phase errors during transport of atoms with a separation
time $\tau_{T}$. Replacing $\varepsilon\left(  t\right)  $\ with
$\varepsilon\left(  x,t\right)  ,$\ we set $\left\langle \varepsilon\left(
x,t\right)  \varepsilon\left(  x^{\prime},t^{\prime}\right)  \right\rangle
=N\left(  \left\vert x-x^{\prime}\right\vert \right)  \int_{-\infty}^{\infty
}S\left(  \omega\right)  e^{i\omega\left(  t-t^{\prime}\right)  }d\omega$ for
transport into or outside of the cavity, where $N\left(  x\right)
=e^{-x^{2}/d^{2}}$, $d=n\lambda/2$ is the distance between two atoms with $n$
integer and $\lambda$ is the wavelength of the counterpropagating laser used
to form 1D optical lattice. The resulting spectral function is $S_{\tau_{T}%
}\left(  \omega\right)  =\int_{-\infty}^{\infty}S\left(  \omega-\nu\right)
\sin^{2}\left[  \left(  \omega-\nu\right)  \tau_{T}/2\right]  \frac
{e^{-\left(  \tau_{T}/4\right)  ^{2}\nu^{2}/2}}{\sqrt{2\pi\left(  4/\tau
_{T}\right)  ^{2}}}d\nu$, which has a suppression of noise with frequencies
$\ll1/\tau_{T}$ by $\left(  \tau_{T}\omega\right)  ^{2}/8$.

\textit{Basic tools.}-----For the logical gate operations, we should introduce
some basic tools---physical controlled-Z (CZ) gate operation and projective
measurements. To perform a collective CZ gate on two atoms \cite{Duan1}, we
reflect a weak coherent light pulse with the so-called odd coherent state from
the cavity, which is resonant with the bare cavity mode and is given in this
form as $\left\vert \alpha_{-}\right\rangle =N_{-}\left(  \left\vert
\alpha\right\rangle -\left\vert -\alpha\right\rangle \right)  $, where $N_{-}$
is normalization constant and $\left\vert \alpha\right\rangle $\ is a coherent
state. Recently, this novel state of light has been generated and
characterized by a non-positive Wigner function experimentally \cite{JS}. For
the case that both atoms are in the state $\left\vert 1\right\rangle $, the
coherent light performed in the limit with $T\gg1/\kappa$ (here $T$ is the
pulse duration and $\kappa$ is the cavity decay rate) is resonantly reflected
by the bare cavity mode with a flipped global phase. For the three other
cases, the effective frequency of the dressed cavity mode will be shifted due
to the atom-cavity coupling, which is described by the Hamiltonian
$H=\hbar\sum_{i=1,2}g_{i}\left(  \left\vert e\right\rangle _{i}\left\langle
0\right\vert a+\left\vert 0\right\rangle _{i}\left\langle e\right\vert
a^{\dagger}\right)  .$ If the coupling rates satisfy $g_{i}\gg\left(
1/T,\kappa,\gamma\right)  $, where $\gamma$\ is the rate of spontaneous decay
of the excited state, then the frequency shift will have a magnitude
comparable with $g_{i}$, so that the incident single-photon pulse will be
reflected by an off-resonant cavity. Hence, both of the shape and global phase
will remain unchanged for the reflected pulse. The net effect of these two
subprocesses is that the reflection of a single-photon pulse from the cavity
actually performs a CZ operation $U_{CZ}=\exp\left(  i\pi\left\vert
11\right\rangle \left\langle 11\right\vert \right)  $ on the two atoms while
leaving the photon state unchanged.

If the input optical pulse is prepared in a weak coherent state $\left\vert
\alpha\right\rangle $, which is reflected following the above analysis from
atom-cavity system, then the projection is obtained after the homodyne
detection of the states of the coherent light as the form%
\begin{equation}
P_{1}=\left\vert 11\right\rangle \left\langle 11\right\vert ;P_{2}=I-P_{1}.
\tag{1}%
\end{equation}

Now we show that by making a little change to the realistic setting one
obtains another projection. Firstly the weak coherent optical pulse enters the
cavity with only atom $1$ inside. After the interaction between atom and
cavity mode, an operation $\exp\left(  i\pi\left\vert 1,\alpha\right\rangle
\left\langle 1,\alpha\right\vert \right)  $ is applied on atom and the optical
pulse. Atom $2$ now is moved into the cavity while $1$ outside, and the pulse
is reflected successively to enter the cavity again, so that the same
operation is applied on atom $2$ and the pulse. After detection, we obtain%
\begin{equation}
P_{3}=\left\vert 00\right\rangle \left\langle 00\right\vert +\left\vert
11\right\rangle \left\langle 11\right\vert ;P_{4}=I-P_{3}. \tag{2}%
\end{equation}

\textit{Logical single qubit operations.-----}The (physical) single qubit
rotation $R_{z}\left(  \alpha\right)  =\exp\left(  -i\alpha\sigma_{z}\right)
$, which can be implemented by RF pulses or the Raman transition applied on
atom 1, has already provided arbitrary logical $z$-rotation, $U_{z}\left(
\alpha\right)  $, i.e., $U_{z}\left(  \alpha\right)  \left\vert 0_{L}%
\right\rangle =e^{-i\alpha}\left\vert 0_{L}\right\rangle $ and $U_{z}\left(
\alpha\right)  \left\vert 1_{L}\right\rangle =e^{i\alpha}\left\vert
1_{L}\right\rangle $.

Then we show another important logical single qubit gate---Hadamard gate.
Consider a system $A$ including atoms $1$ and $2$, on which we want to apply a
Hadamard operation and obtain the outcome state on an ancilla system $B$
including atoms $3$ and $4$ prepared in the state $\left\vert +_{L}%
\right\rangle $\ initially, where $\left\vert \pm_{L}\right\rangle =\left(
\left\vert 0_{L}\right\rangle \pm\left\vert 1_{L}\right\rangle \right)
/\sqrt{2}$. We perform a physical CZ gate on atoms $1$ and $3$, and measure
system $A$ in logical $x$ basis $\left\{  \left\vert +_{L}\right\rangle
,\left\vert -_{L}\right\rangle \right\}  $. If the outcome $\left\vert
-_{L}\right\rangle $ is obtained, we apply $\sigma_{x}\otimes\sigma_{x}$ on
system $B$; else we do nothing. Then $H_{L}=\left(  \left\vert 0_{L}%
\right\rangle \left\langle 0_{L}\right\vert +\left\vert 0_{L}\right\rangle
\left\langle 1_{L}\right\vert +\left\vert 1_{L}\right\rangle \left\langle
0_{L}\right\vert -\left\vert 1_{L}\right\rangle \left\langle 1_{L}\right\vert
\right)  /\sqrt{2}$ is obtained.

Hence, an arbitrary logical single qubit rotation can be implemented with a
sequence of Hadamard operations and $z$-rotations $U=U_{z}\left(
\alpha\right)  H_{L}U_{z}\left(  \beta\right)  H_{L}U_{z}\left(
\varsigma\right)  $.

\textit{Logical single qubit measurements.-----}We can realize logical single
qubit $Z$ measurement of the observable $\sigma_{z}^{L}$ by the sequence of
operations: first, one applies $\sigma_{x}\otimes I$ and then the measurement
$\left\{  P_{1},P_{2}\right\}  $ following by $\sigma_{x}\otimes\sigma_{x}$,
$\left\{  P_{1},P_{2}\right\}  $ again, finally, $I\otimes\sigma_{x}$. The
measurement outcome $\left(  \pi_{1},\pi_{2}\right)  $ with $\pi_{i}$ being
the outcome associated with $P_{i}$, corresponds---in the logical
subspace---to $P_{z,+}^{L}=\left\vert 0_{L}\right\rangle \left\langle
0_{L}\right\vert $; while one obtains $P_{z,-}^{L}=\left\vert 1_{L}%
\right\rangle \left\langle 1_{L}\right\vert $ for the outcome $\left(  \pi
_{2},\pi_{1}\right)  $. Measurements of arbitrary single-qubit observables can
be realized by applying the corresponding basis change.

\textit{Logical Bell-state measurement (BSM).-----}Performing the measurement
$\left\{  P_{3},P_{4}\right\}  $ on atoms $1$ and $3$ belonging to two logical
qubits respectively allows one to distinguish the subspace spanned by
$\left\{  \left\vert \Phi_{L}^{+}\right\rangle ,\left\vert \Phi_{L}%
^{-}\right\rangle \right\}  $ and $\left\{  \left\vert \Psi_{L}^{+}%
\right\rangle ,\left\vert \Psi_{L}^{-}\right\rangle \right\}  $. The
measurement outcomes $\pi_{3}$ and $\pi_{4}$ correspond to $P_{\left\{
\left\vert \Phi_{L}^{+}\right\rangle ,\left\vert \Phi_{L}^{-}\right\rangle
\right\}  }=\left\vert \Phi_{L}^{+}\right\rangle \left\langle \Phi_{L}%
^{+}\right\vert +\left\vert \Phi_{L}^{-}\right\rangle \left\langle \Phi
_{L}^{-}\right\vert $ and $P_{\left\{  \left\vert \Psi_{L}^{+}\right\rangle
,\left\vert \Psi_{L}^{-}\right\rangle \right\}  }=\left\vert \Psi_{L}%
^{+}\right\rangle \left\langle \Psi_{L}^{+}\right\vert +\left\vert \Psi
_{L}^{-}\right\rangle \left\langle \Psi_{L}^{-}\right\vert $, respectively.
More generally, one can obtain non-destructive projections onto subspaces
spanned by two arbitrary Bell states using additional logical single qubit
unitary operations which allow one to permute Bell states. For instance, the
application $H_{L}\otimes H_{L}$ consequently before and after the measurement
$P_{\left\{  \left\vert \Phi_{L}^{+}\right\rangle ,\left\vert \Phi_{L}%
^{-}\right\rangle \right\}  }$ corresponds to $P_{\left\{  \left\vert \Phi
_{L}^{+}\right\rangle ,\left\vert \Psi_{L}^{+}\right\rangle \right\}  }$.
Obviously, using these non-destructive projections, we can achieve a full
logical BSM.

\textit{Two-qubit gate.-----}A logical CZ gate described by $U_{CZ}%
^{L}=diag\left(  1,1,1,-1\right)  $ in the logical basis, can be realized
shown in Fig. 1a via atoms in a cavity by performing a physical CZ operation
on atoms $1$ and $3$ belonging to two logical qubits respectively.

Now we analyze the fidelity of the logical CZ gate under the influence of some
practical sources of noise. For the initial state of the system $\left\vert
\Psi_{in}\right\rangle =\sum_{m,n=0,1}\epsilon_{mn}\left\vert m_{L}%
\right\rangle \left\vert n_{L}\right\rangle \left\vert \varphi\right\rangle
_{in}$, $\left\vert \varphi_{in}\right\rangle \propto\left\{  \exp\left[
\alpha\int_{0}^{T}f_{in}\left(  t\right)  a_{in}^{\dagger}\left(  t\right)
dt\right]  -\exp\left[  -\alpha\int_{0}^{T}f_{in}\left(  t\right)
a_{in}^{\dagger}\left(  t\right)  dt\right]  \right\}  \left\vert
\text{vac}\right\rangle $ is the state of the input coherent optical pulse
with a normalized shape function $f_{in}\left(  t\right)  $, where $\left\vert
\text{vac}\right\rangle $ denotes the vacuum state and $a_{in}^{\dagger
}\left(  t\right)  $ is the one-dimensional optical field operator with the
commutation relation $\left[  a_{in}\left(  t\right)  ,a_{in}^{\dagger}\left(
t^{\prime}\right)  \right]  =\delta\left(  t-t^{\prime}\right)  $. The cavity
mode $a_{c}$ is driven by the input field $a_{in}\left(  t\right)  $ through
the Langevin equation $\dot{a}_{c}=-i[a_{c},H]-\left(  \kappa/2\right)
a_{c}-\sqrt{\kappa}a_{in}\left(  t\right)  $. The output field $a_{out}\left(
t\right)  $ of the cavity is connected with the input through the input-output
relation $a_{out}\left(  t\right)  =a_{in}\left(  t\right)  +\sqrt{\kappa
}a_{c}$. The output state of the whole system can be written as $\left\vert
\Psi_{out}\right\rangle =\sum_{m,n=0,1}e^{i\theta_{mn}}\epsilon_{mn}^{\prime
}\left\vert m_{L}\right\rangle \left\vert n_{L}\right\rangle \left\vert
\varphi_{out}\right\rangle _{mn}$, where the output state of the coherent
light $\left\vert \varphi_{out}\right\rangle _{mn}$ corresponds to the atomic
component $\left\vert m_{L}\right\rangle \left\vert n_{L}\right\rangle $ with
a shape $f_{mn}^{out}\left(  t\right)  $ and amplitude $\alpha_{mn}^{\prime}$.
In general, the amplitude $\alpha_{mn}^{\prime}$ (for $m,n\neq1$) is different
from $\alpha$ because of the effect of the atomic spontaneous emission
loss-----the fundamental source of photon loss in cavity can be quantified by
the photon loss parameter $\eta=1-\left\vert \alpha^{\prime}\right\vert
^{2}/\left\vert \alpha\right\vert ^{2}\propto\kappa\gamma/g_{o}^{2}$ through
the numerical simulations. Ideally, the output state $\left\vert \Psi
_{out}^{id}\right\rangle $ would have the unchanged amplitude $\alpha$ and
shape functions $f_{11}^{out}\left(  t\right)  =-f_{in}\left(  t\right)  $ and
$f_{mn}^{out}\left(  t\right)  =f_{in}\left(  t\right)  $ (for $m,n\neq1$).
Then the fidelity can be defined as $F\equiv\left\vert \left\langle \Psi
_{out}^{id}\right\vert \left\vert \Psi_{out}\right\rangle \right\vert ^{2},$
which decreases with the mean photon number $\langle n\rangle=\left\vert
\alpha\right\vert ^{2}$.

We investigate the fidelity under typical experimental configurations and it
is shown in Fig 2(a) as a function of the mean photon number of the input
state for the realistic parameters $\left(  g_{o},\kappa,\gamma\right)
/2\pi=\left(  27,2.4,2.6\right)  $ MHz \cite{Sauer}. We obtain a high fidelity
up to $0.99$ for these parameters and the coherent input pulse with a
remarkable amplitude $\alpha=1.26$. Furthermore, $F$ is insensitive to the
variation of the coupling rate caused by fluctuations in atomic position, and
$\delta F$ describing change of the fidelity is about $10^{-2}$ for $g$
varying to $g/2$.

\begin{figure}[ptb]
\includegraphics[width=8cm,height=4cm]{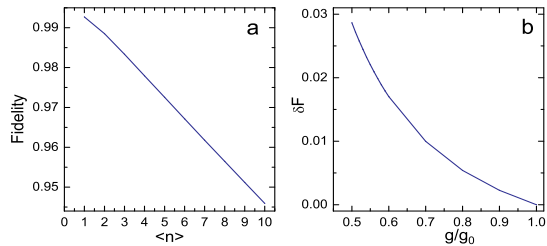}\newline\caption{(a) The
fidelity of the logical CZ gate versus the mean photon number of the coherent
optical pulse with the pulse duration $T=200/\kappa,$ and (b) it changes with
$g/g_{o}$. We have assumed a Gaussian shape for the input pulse with $f\left(
t\right)  \propto\exp\left[  -\left(  t-T/2\right)  ^{2}/\left(  T/5\right)
^{2}\right]  .$ Here we choose the realistic parameters $\left(  g_{o}%
,\kappa,\gamma\right)  /2\pi=\left(  27,2.4,2.6\right)  $ MHz.}%
\end{figure}

The above scheme can also be extended to perform logical CNOT gate---in
principle between two logical qubits represented by remote atoms trapped in
different cavities at arbitrary distance since the (physical) CZ gate can be
implemented between two atoms belonging to different logical qubits in
separated cavities \cite{Duan1,Xiao}.

The entangled state $\left\vert \Phi_{L}^{+}\right\rangle _{AB}$ is used to
generate the logical four-qubit state that corresponds to CNOT gate. We use
notation $A$, $A^{^{\prime}}$, $B$, $B^{^{\prime}}$ to refer to different
atoms, where $A$ and $A^{^{\prime}}$ referring atoms trapped in cavity $1$
belong to one party, while $B$ referring atoms trapped in cavity $1$ and
$B^{^{\prime}}$ in cavity $2,$ belong to another separated party. We prepare
two ancilla logical qubits $A^{^{\prime}}$ and $B^{^{\prime}}$ in the states
$\left\vert +_{L}\right\rangle _{A^{\prime}}$ and $\left\vert 0_{L}%
\right\rangle _{B^{\prime}}$, i.e. the initial state is $\left\vert
\zeta\right\rangle =\left\vert +_{L}\right\rangle _{A^{\prime}}\left\vert
\Phi_{L}^{+}\right\rangle _{AB}\left\vert 0_{L}\right\rangle _{B^{\prime}}$.
The following sequence of operations with indicated measurement outcomes
generates the desired state: $P_{\left\{  \left\vert \Phi_{L}^{+}\right\rangle
,\left\vert \Phi_{L}^{-}\right\rangle \right\}  }^{AA^{\prime}}P_{\left\{
\left\vert \Phi_{L}^{+}\right\rangle ,\left\vert \Psi_{L}^{+}\right\rangle
\right\}  }^{BB^{\prime}}\left\vert \zeta\right\rangle =\left(  \left\vert
0_{L}0_{L}\right\rangle _{AA^{\prime}}\left\vert \Phi_{L}^{+}\right\rangle
_{BB^{\prime}}+\left\vert 1_{L}1_{L}\right\rangle _{AA^{\prime}}\left\vert
\Psi_{L}^{+}\right\rangle _{BB^{\prime}}\right)  /\sqrt{2}\equiv\left\vert
\Xi\right\rangle .$

Given two additional logical qubits in an arbitrary state $\rho_{A^{\prime
\prime}B^{\prime\prime}}$, where $A^{\prime\prime}$ and $B^{\prime\prime}$
refer atoms trapped in cavity $2$ and $1$, respectively, one can use the state
$\left\vert \Xi\right\rangle $ together with logical BSM, to implement a
logical CNOT operation on $\rho_{A^{\prime\prime}B^{\prime\prime}}$ and obtain
the outcome state on system $A^{\prime}B^{\prime}$ following the procedure
shown in \cite{Gott}. This is achieved by measuring systems $A^{\prime\prime
}A$ and $B^{\prime\prime}B$\ in the logical Bell basis $\left\vert \psi
_{i_{1},i_{2}}\right\rangle =I\otimes\sigma_{i_{1},i_{2}}^{L}\left\vert
\Phi_{L}^{+}\right\rangle $, where $\sigma_{i_{1},i_{2}}^{L}$ is one of Pauli
operators. If the outcome for $A^{\prime\prime}A$\ is $\left\vert \psi
_{i_{1},i_{2}}\right\rangle ,$\ we apply $\sigma_{i_{1},i_{2}}^{L}$\ on
$A^{\prime}$ and proceed analogously with $B^{\prime\prime}B.$ One can readily
see that the resulting operation on $A^{\prime}B^{\prime}$ after the procedure
will be $U_{CNOT}$ or $U_{CNOT}^{+}$ with the same probability$.$ Since
$U_{CNOT}=U_{CNOT}^{+}$, we obtain a deterministic implementation of logical
CNOT gate, and then atoms $A^{\prime}$, $B^{\prime}$ are in the state
$U_{CNOT}\rho_{A^{\prime\prime}B^{\prime\prime}}U_{CNOT}^{+}$.

\textit{Leakage error detection.-----}A method is presented to detect leakage
errors, in which the state within the logical subspace $\left\{  \left\vert
0_{L}\right\rangle ,\left\vert 1_{L}\right\rangle \right\}  $ is not altered.
Consider a system $A$ in some pure state $\left\vert \varphi\right\rangle
=\left\vert \chi\right\rangle +\left\vert \chi^{\perp}\right\rangle $, where
$\left\vert \chi\right\rangle $\ is a state belonging to the logical subspace
spanned by $\left\{  \left\vert 0_{L}\right\rangle ,\left\vert 1_{L}%
\right\rangle \right\}  $, while $\left\vert \chi^{\perp}\right\rangle
$\ belongs to the orthogonal subspace $\left\{  \left\vert 2_{L}\right\rangle
=\left\vert 00\right\rangle ,\left\vert 3_{L}\right\rangle =\left\vert
11\right\rangle \right\}  $\ and corresponds to leakage error. An ancilla
system $B$ is prepared in $\left\vert +_{L}\right\rangle ,$ and then the
measurement $\left\{  P_{3},P_{4}\right\}  $ is performed on atoms $2$ and $4$
and then on $1$ and $4$. If the same outcomes in two measurements are
obtained, i.e. $\left(  \pi_{3},\pi_{3}\right)  $ and $\left(  \pi_{4},\pi
_{4}\right)  $, that means the initial system was outside of the logical
subspace. In these cases we conclude that leakage error occurred. For the
different outcomes $\left(  \pi_{3},\pi_{4}\right)  $ or $\left(  \pi_{4}%
,\pi_{3}\right)  $, we perform a logical CNOT operation on systems $A$ and
$B$, then the state of system $A$ is given by $\left\vert \chi\right\rangle $.
Hence, this procedure always provides a conclusive leakage error detection.

\textit{Feasibility of the proposal.-----}No particularly demanding
assumptions have been made for experimental parameters. The relevant cavity
QED parameters for our system are assumed as $g_{o}^{2}/\kappa\gamma=51\gg1,$
placing our system well into the strongly coupled regime. The cavity consists
of two $1$-mm-diam mirrors with $10$ cm radii of curvature separated by $75$
$\mu$m \cite{Sauer} assuming the wavelength of the cavity mode is $\sim780$ nm
(the rubidium $D2$ line). The distance between two atoms $d$ in an optical
lattice has a magnitude of $10$ $\mu$m, which is larger than the waist $\sim5$
$\mu$m to leave only one atom inside the cavity and its neighbor atoms outside
for the logical gate operations. The evolution of the states of two atoms is
accomplished in the duration of the single-photon pulse $T\sim200/\kappa=13$
$\mu$s. The maximum velocity of the atoms in the transverse optical lattices
is $30$ cm/s and the maximum acceleration imparted is $1.5g$. Moving the
proper atoms into and outside of the cavity is accomplished in the time
$\tau_{T}$ of $100$ $\mu$s. The gate preformation and transport of atoms can
be accomplished within the coherent time (dephasing) of atoms with a magnitude
of milliseconds \cite{Ja,Bonn}. Hence, our scheme fits well the status of
current experimental technology.

\textit{Summary.-----}We have proposed a scheme for deterministic quantum
gates acting on neutral atoms in DFS which from the beginning immunizes our
logical qubits against the dominant source of decoherence-----dephasing. The
efficiency of this scheme is characterized through exact numerical simulations
that incorporate various sources of experiment noise and these results
demonstrate the practicality by way of current experimental technology. Some
processes proposed here such as full BSM and unitary operations based on
teleportation may also find applications in quantum communication and metrology.

\begin{acknowledgments}
We thank P. Zoller for critical remarks, and W. D\"{u}r, H. Briegel, L.M.
Duan, B. Wang, Z.B. Chen and S.Chen for stimulating discussions. PX was
supported by the Austrian Academy of Science. YFX was funded by the Knowledge
Innovation Project of Chinese Academy of Sciences.
\end{acknowledgments}

\end{document}